\documentstyle[12pt,./epsf,a4wide]{article}
\def \refs {\begingroup \frenchspacing
    \parskip = 0.2 \baselineskip \parindent = 0 pt
    \everypar = {\hangindent = 20.0 pt \hangafter = 1}}
\def \endrefs {\par \endgroup}
\def \la {\mathrel{\vcenter
     {\offinterlineskip \hbox{$<$}\hbox{$\sim$}}}}
\def \ga {\mathrel{\vcenter
     {\offinterlineskip \hbox{$>$}\hbox{$\sim$}}}}
\def\ave#1{\langle #1 \rangle}
\def\eck#1{\left\lbrack #1 \right\rbrack}
\def\rund#1{\left( #1 \right)}

\font\top=cmbx12 scaled \magstep2
\begin{document}
\title{Ledoux-convection in protoneutron stars --
       a clue to supernova nucleosynthesis~?}
\author{Wolfgang Keil, H.-Thomas Janka, and Ewald M\"uller\\
Max-Planck-Institut f\"ur Astrophysik\\
Karl-Schwarzschild-Str.~1, D-85740 Garching, Germany
}
\maketitle
 
\vskip 1.0 truecm
\centerline{Accepted for publication in 
{\it The Astrophysical Journal, Letters}}
\vskip 2.0 truecm
 
\centerline{ABSTRACT}
\vskip 0.5 truecm
Two-dimensional (2D) hydrodynamical simulations of the deleptonization
of a newly formed neutron star (NS) were performed.
Driven by negative lepton fraction and entropy gradients,
convection starts near the neutrinosphere about 20--30~ms
after core bounce, but moves deeper into the protoneutron star (PNS),
and after about one second the whole PNS is convective.
The deleptonization of the star proceeds much faster than in the
corresponding spherically symmetrical model because the lepton flux
and the neutrino ($\nu$) luminosities
increase by up to a factor of two. The convection below the
neutrinosphere raises the neutrinospheric temperatures and mean
energies of the emitted $\nu$'s by 10--20\%. This can have
important implications for the supernova (SN) explosion mechanism and
changes the detectable $\nu$ signal from the Kelvin-Helmholtz
cooling of the PNS. In particular, the enhanced $\nu_e$ flux
relative to the $\bar\nu_e$ flux during the early post-bounce
evolution might solve the overproduction problem of certain elements
in the neutrino-heated ejecta in models of type-II SN explosions.

\bigskip
\noindent
{\it Subject headings:} supernovae: general -- stars: neutron --
elementary particles: neutrinos -- turbulence -- convection --
hydrodynamics


\section{Introduction}

Convection in the newly formed NS might play an important
role to explain the explosion of a massive star in a type-II SN.
Epstein (1979) pointed out that not only entropy, $S$, inversions
but also zones in the post-collapse core where the lepton fraction,
$Y_l$, decreases with
increasing radius tend to be unstable against Ledoux convection.
Negative $S$ and/or $Y_l$
gradients in the neutrinospheric region and in the layers between the
nascent NS and the weakening prompt shock front were realized in
a variety of post-bounce SN models by Burrows \& Lattimer (1988),
and after shock stagnation
in computations by Hillebrandt (1987) and more recently
by Bruenn (1993), Bruenn \& Mezzacappa (1994),
and Bruenn et al.~(1995). Despite different equations of states
(EOS), $\nu$ opacities, and $\nu$ transport methods,
the development of negative $Y_l$ and $S$ gradients is common
in these simulations and can also be found in PNS cooling models of
Burrows \& Lattimer (1987), Keil \& Janka (1995), and
Sumiyoshi et al.~(1995).

Convection above the neutrinosphere but below the
neu\-tri\-no-\-heat\-ed
region can hardly be a direct help for the explosion
(Bethe et al.~1987, Bruenn et al.~1995),
whereas convectively enhanced lepton number and energy
transport inside the neutrinosphere
raise the $\nu$ luminosities and can definitely support
neutrino-energized SN explosions (Bethe et al.~1987).
In this context, Burrows (1987) and Burrows \& Lattimer (1988)
have discussed
entropy-driven convection in the PNS on the basis
of 1D, general relativistic (GR) simulations of the first
second of the evolution of a hot, $1.4\,M_{\odot}$ PNS.
Their calculations were done with a Henyey-like code using a
mixing-length scheme for convective energy and lepton transport.
Recent 2D models (Herant et al.~1994, Burrows et al.~1995,
Janka \& M\"uller 1996 and references therein)
confirmed the possibility that convective
processes can occur in the surface region of the PNS immediately
after shock stagnation (``prompt convection'') for a period of at least
several 10~ms. These models, however, have been evolved only over
rather short times or with insufficient numerical resolution in the
PNS or with a spherically symmetrical
description of the core of the PNS that was in some
cases even replaced by an inner boundary condition.

Mayle \& Wilson (1988) and Wilson \& Mayle (1988, 1993)
demonstrated that convection in the nascent NS
can be a crucial ingredient that leads to successful delayed explosions.
With the high-density EOS and treatment of the $\nu$
transport used by the Livermore group, however, negative
gradients of $Y_l$ tend to be stabilized by positive
$S$ gradients (see, e.g., Wilson \& Mayle 1989). Therefore
they claim doubly diffusive neutron finger convection to
be more important than Ledoux convection. Doubts about the presence
of doubly diffusive instabilities, on the other hand, were recently
raised by Bruenn \& Dineva (1996).
Bruenn \& Mezzacappa (1994) and Bruenn et al.~(1995) also come to
a negative conclusion about the relevance of prompt convection
in the neutrinospheric region. Although
their post-bounce models show unstable $S$ and $Y_l$
stratifications, the mixing-length approach in their 1D
simulations predicts convective activity inside and around the
neutrinosphere to be present only for 10--30~ms after bounce and to
have no significant impact on the $\nu$ fluxes and spectra when
an elaborate multi-group flux-limited diffusion method is used for
the $\nu$ transfer. Such conclusions seem to be supported by
recent 2D simulations of Mezzacappa et al.~(1996). These 2D models,
however, still suffer from the use of an inner boundary condition
at a fixed radius of 20--30~km, a simplified treatment of 
neutrino-matter interactions, and imposed neutrino fluxes and
spectra from spherically symmetrical models.

From these differing and partly contradictory
results it is evident that the question whether, where,
when, and how long convection occurs below the neutrinosphere
seems to be a matter of the EOS, of the core structure
of the progenitor star, of the shock properties and propagation,
and of the $\nu$ opacities and the $\nu$ transport description.
In this Letter we compare 1D simulations with the first
self-consistent 2D models that follow the
evolution of the newly formed NS for more than a second,
taking into account the GR gravitational
potential and making use of a flux-limited equilibrium diffusion scheme
that describes the transport of $\nu_e$, $\bar\nu_e$, and $\nu_x$
(sum of $\nu_{\mu}$, $\bar\nu_{\mu}$, $\nu_{\tau}$, and $\bar\nu_{\tau}$)
and is very good at high optical depths but only approximative
near the PNS surface (Keil \& Janka 1995).
Our simulations demonstrate that Ledoux convection may continue
in the PNS for a long time and can involve
the whole star after about one second.


\section{Numerical implementation\label{sec:numerics}}

The simulations were performed with the explicit Eulerian
hydrodynamics code {\it Prometheus} (Fryxell, M\"uller, \& Arnett~1989)
that employs a Riemann-solver and is based on the Piecewise
Parabolic Method (PPM) of Colella \& Woodward (1984). A moving grid
with 100 nonequidistant radial zones (initial outer radius
$\sim 60\,{\rm km}$, final radius $\sim 20\,{\rm km}$)
and with up to 60 angular zones was used, corresponding
to a radial resolution of a few 100~m
($\la 1\,{\rm km}$ near the center) and a maximum
angular resolution of $1.5^{\circ}$.
In the angular direction, periodic boundary
conditions were imposed at $\pm 45^{\circ}$ above and below the
equatorial plane. The stellar surface was treated as an
open boundary where the velocity was calculated from the velocity
in the outermost grid zone, the density profile was extrapolated
according to a time-variable power law, and the corresponding
pressure was determined from the condition of hydrostatic
equilibrium. The {\it Prometheus} code was extended for the use of
different time steps and angular resolutions in different
regions of the star. Due to the extremely restrictive
Courant-Friedrichs-Lewy (CFL) condition
for the hydrodynamics, the implicit $\nu$ transport was computed
typically with 10 times larger time steps than the smallest
hydrodynamics time step on the grid
($\sim 10^{-7}\,{\rm s}$) (Keil 1996).

Our simulations were started with the $\sim 1.1\,M_{\odot}$
(baryonic mass) central, dense part
($\rho \ga 10^{11}\,{\rm g/cm}^3$) of the collapsed core of a
$15\,M_{\odot}$ progenitor star (Woosley et al.~1988) that was
computed to a time of about 25~ms after core bounce (i.e., a
few ms after the stagnation of the prompt shock) by Bruenn (1993).
Accretion was not considered but additional matter could be
advected onto the grid through the open outer boundary.
In the 2D run, Newtonian asphericity corrections
were added to the spherically symmetrical GR
gravitational potential: $\Phi_{\rm 2D} \equiv \Phi_{\rm 1D}^{\rm GR}
+ \rund{\Phi_{\rm 2D}^{\rm N}-\Phi_{\rm 1D}^{\rm N}}$. This should
be a sufficiently good approximation because
convective motions produce only local and minor deviations of the
mass distribution from spherical symmetry. Using the GR
potential ensured that transients due to the mapping of Bruenn's
(1993) relativistic 1D results to our code were very small.
When starting our 2D simulation, the radial velocity
(under conservation of the local specific total energy)
was randomly perturbed in the whole PNS with an
amplitude of 0.1\%. The thermodynamics of the NS
medium was described by the EOS of Lattimer \& Swesty (1991) which
yields a physically reasonable description of nuclear matter below
about twice nuclear density and is thus suitable to describe the
interior of the considered low-mass NS ($M_{\rm ns}\la 1.2\,M_{\odot}$).

The $\nu$ transport was carried out in radial direction for
every angular zone of the finest angular grid. Angular transport
of neutrinos was neglected. This underestimates the
ability of moving buoyant fluid elements to exchange lepton number
and energy with their surroundings and is only correct if
radial radiative and convective transport are faster. Moreover,
$\nu$ shear viscosity was disregarded.
For $\nu$-$n,\,p$ scattering and 3 flavors of nondegenerate
$\nu$ and $\bar\nu$ in local thermal equilibrium with the
matter one estimates at low densities
($\rho \la 10^{14}\,{\rm g/cm^3}$) a dynamic shear viscosity of
$\eta_{\nu}^{\rm nd} \sim 1.2\cdot 10^{22}T_{10}^2/\rho_{14}\,
{\rm g\,cm^{-1}s^{-1}}$ (van den Horn \& van Weert 1981), and in
degenerate nuclear matter
$\eta_{\nu}^{\rm d}  \sim 6.7\cdot 10^{22}f(Y_p)T_{10}/\rho_{14}^{1/3}
\,{\rm g\,cm^{-1}s^{-1}}$ (Thompson \& Duncan 1993) when
$T_{10}\equiv T/(10\,{\rm MeV})$ and
$\rho_{14}\equiv \rho/(10^{14}\,{\rm g/cm^3})$.
The expression $f(Y_p)\approx 0.63\,...\,1$ is a function
of the proton fraction $Y_p$. With convective
velocities $v_{\rm c}\sim 10^8\,{\rm cm/s}$ and length scales of
convective
mixing $l_{\rm c}\sim 10^5\,{\rm cm}$, one obtains Reynolds numbers
${\cal R}_{\nu} = v_{\rm c}l_{\rm c}\rho/\eta_{\nu}\sim
10^4...\,10^5$ for typical temperatures and densities in the PNS,
and corresponding viscous damping timescales of convective motions
of the order of
$t_{\nu} \sim {1\over 2}\,l_{\rm c}^2\rho/\eta_{\nu}\approx 5$--50~s.
Neutrino viscosity is not extremely important on
timescales of $\sim 1\,{\rm s}$ and should not be able to
suppress turbulent convection (Thompson \& Duncan 1993,
Burrows \& Lattimer 1988).

The effective numerical viscosity of the
PPM scheme is estimated from Eq.~(3.6) in Porter \& Woodward (1994)
to be $\eta_{\rm n}/\rho \sim
v_{\rm c}l_{\rm c}10^y/(2\pi)^2$, which depends on the grid resolution
via the parameter $y$ defined as the ordinate in
Fig.~1 of Porter \& Woodward (1994). For structures of size
$l_{\rm c}/(10^5\,{\rm cm})\equiv l_{{\rm c},5}\ga 1$ that are
typically resolved by about 10 zones,
and for flows with advective Courant numbers $C_{\rm a}\ga 0.08$,
Fig.~1 of Porter \& Woodward (1994) leads to a dynamic shear
viscosity of $\eta_{\rm n}\la 5\cdot 10^{23}\rho_{14}l_{{\rm c},5}
v_{{\rm c},8}\,{\rm g\,cm^{-1}s^{-1}}$ when
$v_{{\rm c},8}\equiv v_{\rm c}/(10^8\,{\rm cm/s})$. The corresponding
Reynolds numbers are ${\cal R}_{\rm n}\sim (2\pi)^2 10^{-y}\ga 2000$,
and Eq.~(3.5) of Porter \& Woodward (1994) yields for the viscous
damping timescale $t_{\rm n}\sim {1\over 2}
\cdot 10^{-y}l_{\rm c}/v_{\rm c}\ga 25\,l_{{\rm c},5}/v_{{\rm c},8}\,
{\rm ms}$. This means that small structures represented
by only a few grid cells will be affected by the numerical viscosity,
but damping timescales for fluid motions on scales
$l_{\rm c}\ga 10^5\,{\rm cm}$
are still a factor ${1\over 2}\cdot 10^{-y}\ga 25$ longer than
the overturn timescales $t_{\rm o} \sim l_{\rm c}/v_{\rm c}\sim
{\rm few~ms}$.


\section{Results\label{sec:results}}

\begin{figure}
\centering\leavevmode
\epsfxsize=16cm
\epsfbox{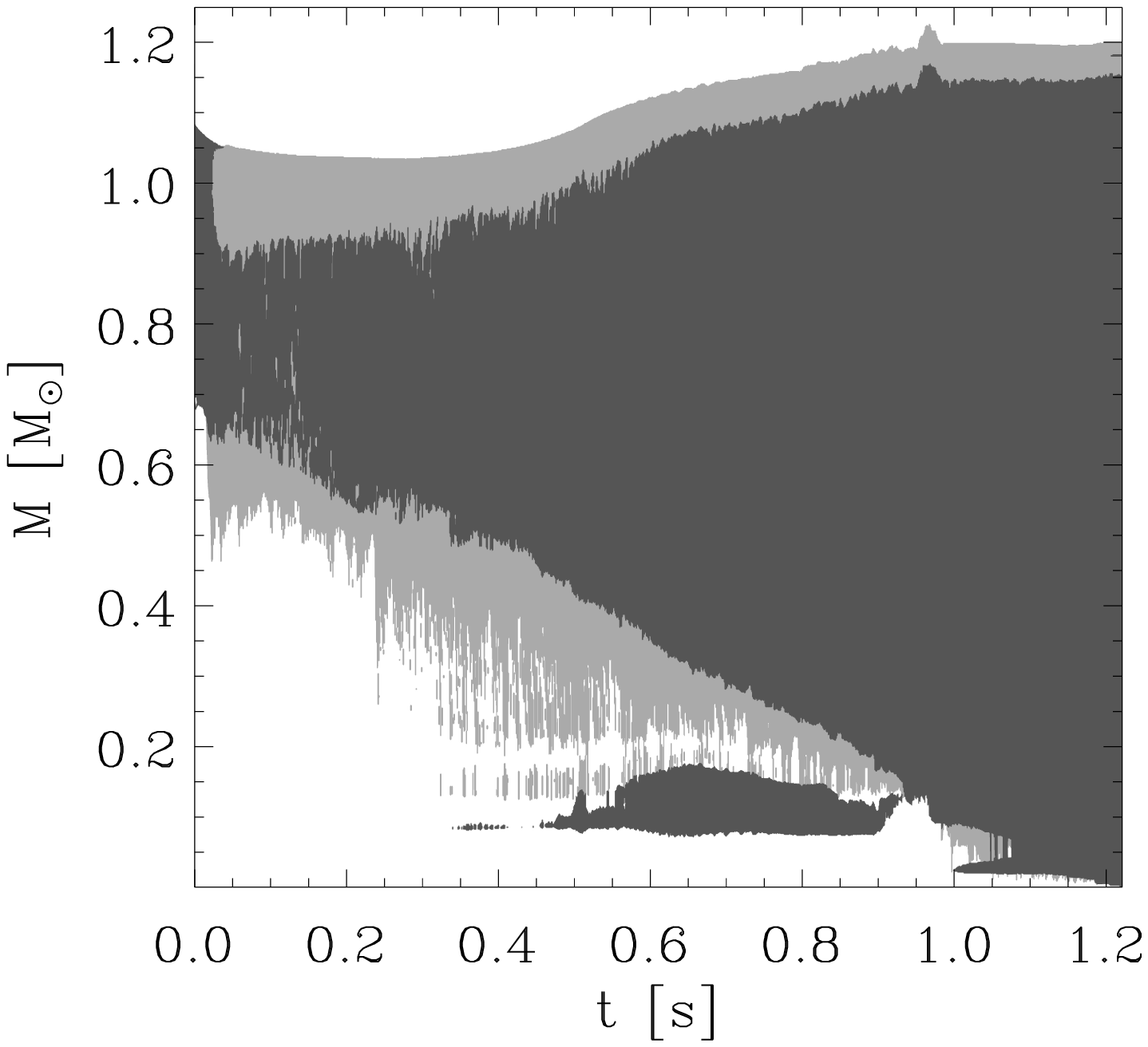}
\caption[]{
}
\label{fig:conv-map}
\end{figure}

\begin{figure}
\centering\leavevmode
\epsfxsize=16cm
\epsfbox{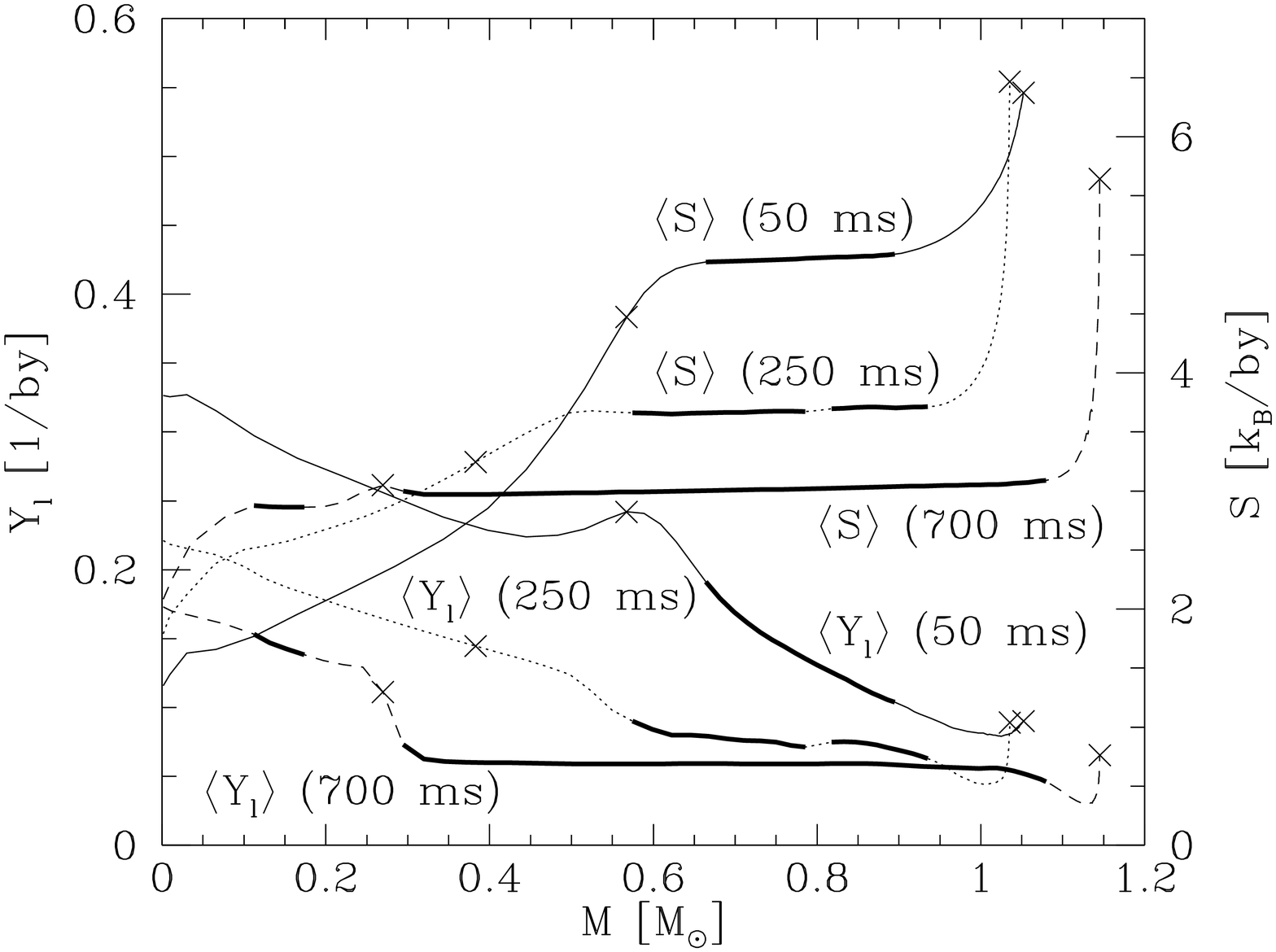}
\caption[]{ 
}
\label{fig:sy-cuts}
\end{figure}

Convection can be driven by a radial gradient of the entropy
per nucleon $S$ and/or by a gradient of the lepton number per
baryon $Y_l$ (Epstein 1979) where $Y_l$ includes contributions
from $e^-$ and $e^+$ and from $\nu_e$ and $\bar\nu_e$ if
the latter are in equilibrium with the matter. Convective
instability in the Ledoux approximation sets in when
%
\begin{equation}
{\cal C}_{\rm L}(r)\,\equiv\,
\rund{{\partial \rho\over\partial S}}_{\!\! P,Y_l}
{{\rm d}S\over{\rm d}r} +
\rund{{\partial \rho\over\partial Y_l}}_{\!\! P,S}
{{\rm d}Y_l\over{\rm d}r} \,>\, 0 \ .
\label{eq:ledoux}
\end{equation}
%
Initially, this criterion is fulfilled between $\sim 0.7\,M_{\odot}$
and $\sim 1.1\,M_{\odot}$ (black area in
Fig.~\ref{fig:conv-map}; see also Bruenn et al.~1995) and convective
activity develops within $\sim 10\,{\rm ms}$ after the start of
the 2D simulation. About 30~ms later the outer layers become
convectively stable which is in agreement with
Bruenn \& Mezzacappa (1994). In our 2D simulation, however, the
convectively unstable region
retreats to mass shells $\la 0.9\,M_{\odot}$ and its inner edge
moves deeper into the neutrino-opaque interior of the
star, following a steeply negative lepton
gradient that is advanced towards the stellar center by the
convectively enhanced deleptonization of the outer layers
(Figs.~\ref{fig:conv-map}~and~\ref{fig:sy-cuts}). Note that
the black area in Fig.~\ref{fig:conv-map} and the thick solid
lines in Fig.~\ref{fig:sy-cuts} mark not only
those regions in the star which are convectively unstable but also
those {\it which are only marginally
stable} according to the Ledoux criterion of Eq.~(\ref{eq:ledoux})
for angle-averaged $S$ and $Y_l$,
i.e., regions where ${\cal C}_{\rm L}(r) \ge
a\cdot\max_r(|{\cal C}_{\rm L}(r)|)$
with $a = 0.05$ holds. For $a\la 0.1$ the accepted region varies
only little with $a$ and is always embedded by the grey-shaded
area where $|v_{\theta}| > 10^7\,{\rm cm/s}$.
Yet, only sporadically and randomly appearing patches
in the convective layer fulfill Eq.~(\ref{eq:ledoux}) rigorously.
Figure~\ref{fig:sy-cuts} shows that the black region
in Fig.~\ref{fig:conv-map} coincides with the layers
where convective mixing flattens the $S$ and $Y_l$ gradients.

\begin{figure*}
 \begin{tabular}{cc}
 \hskip -1.0 truecm
 \epsfxsize=10.25cm \epsfclipon \epsffile{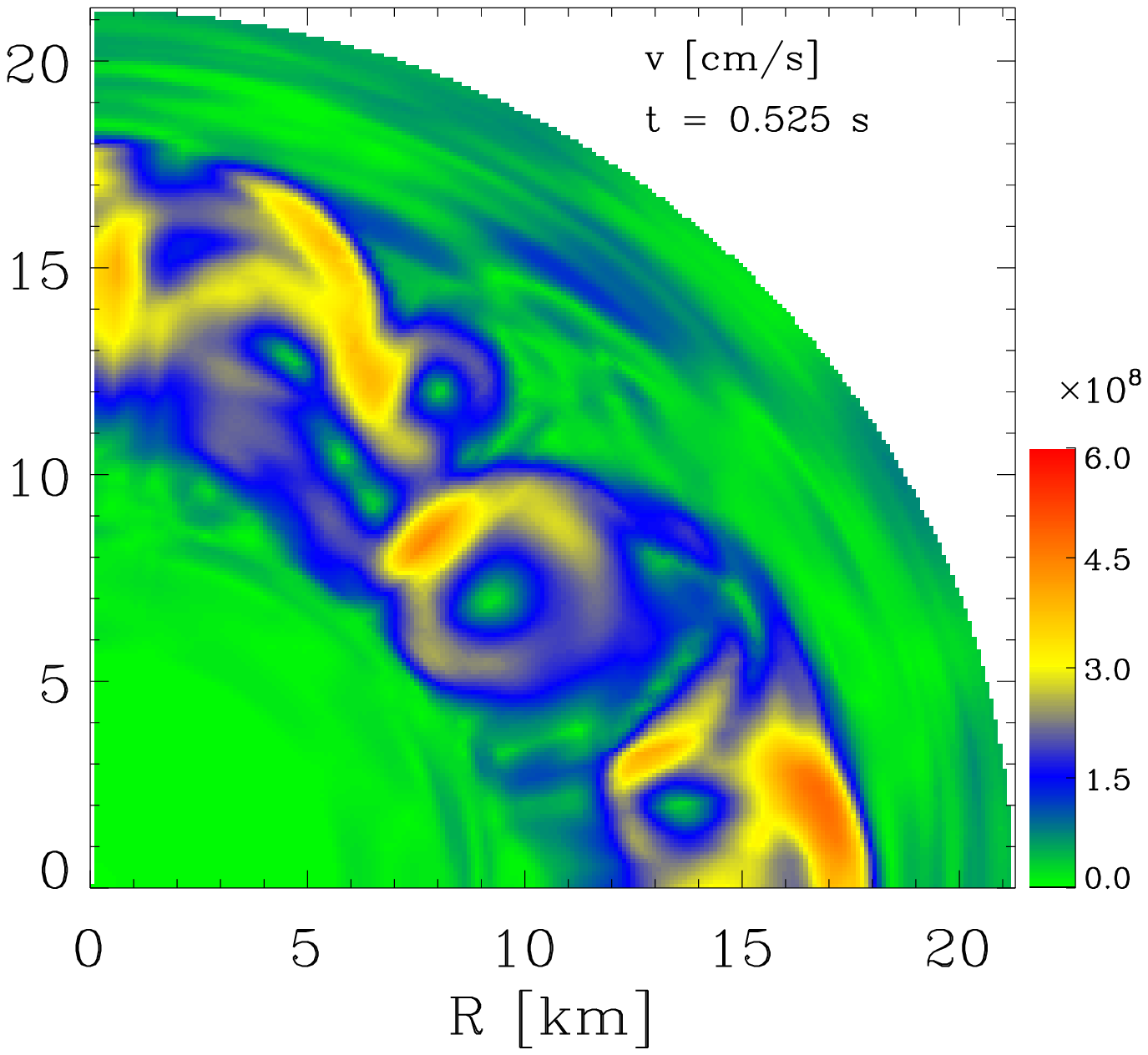}
 \put(-2.0,7.5){{\Large \bf a}} &
 \hskip -1.7 truecm
 \epsfxsize=10.25cm \epsfclipon \epsffile{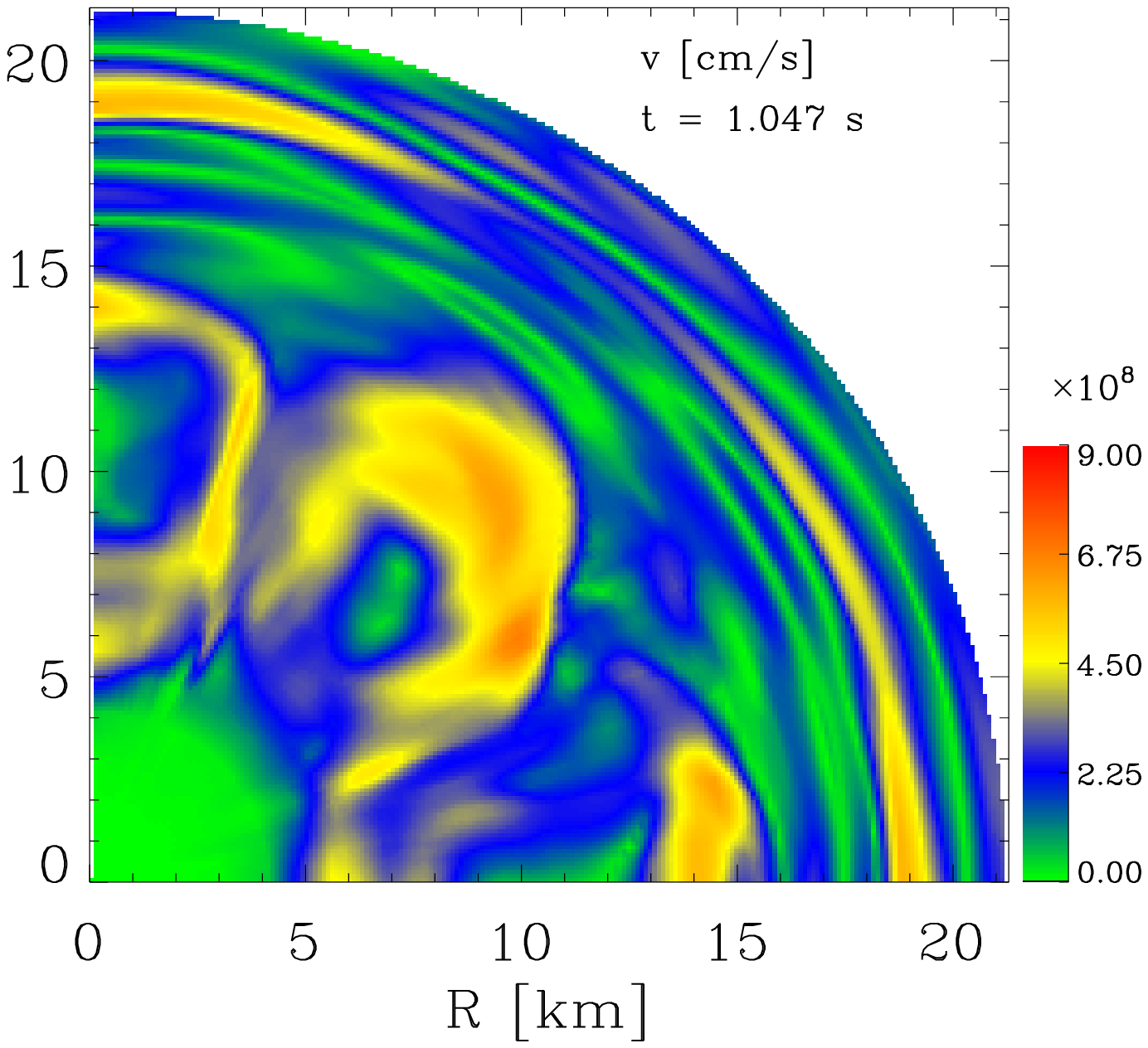}
 \put(-2.0,7.5){{\Large \bf b}} \\
 \hskip -1.0 truecm
 \epsfxsize=10.25cm \epsfclipon \epsffile{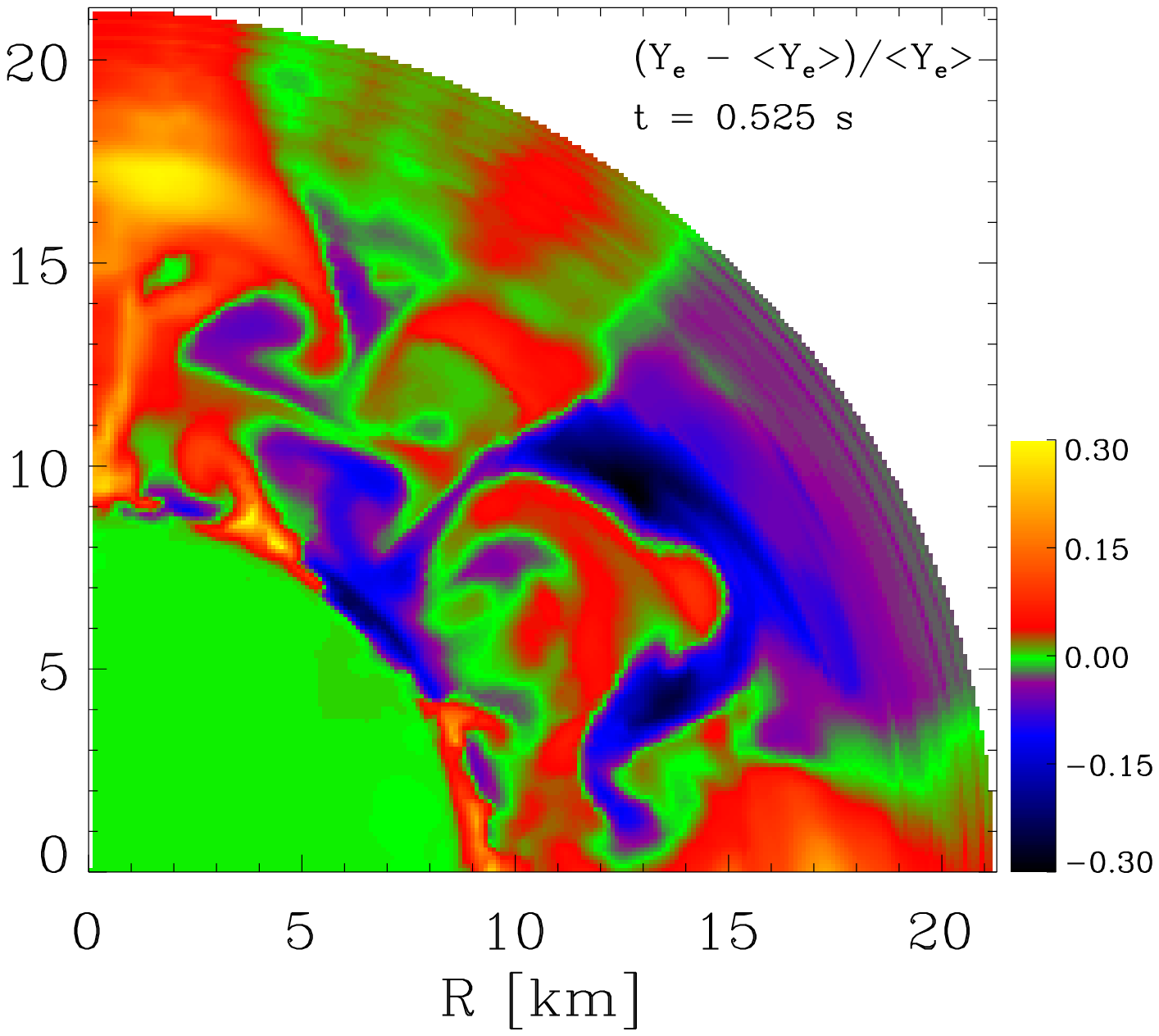}
 \put(-2.0,7.5){{\Large \bf c}} &
 \hskip -1.7 truecm
 \epsfxsize=10.25cm \epsfclipon \epsffile{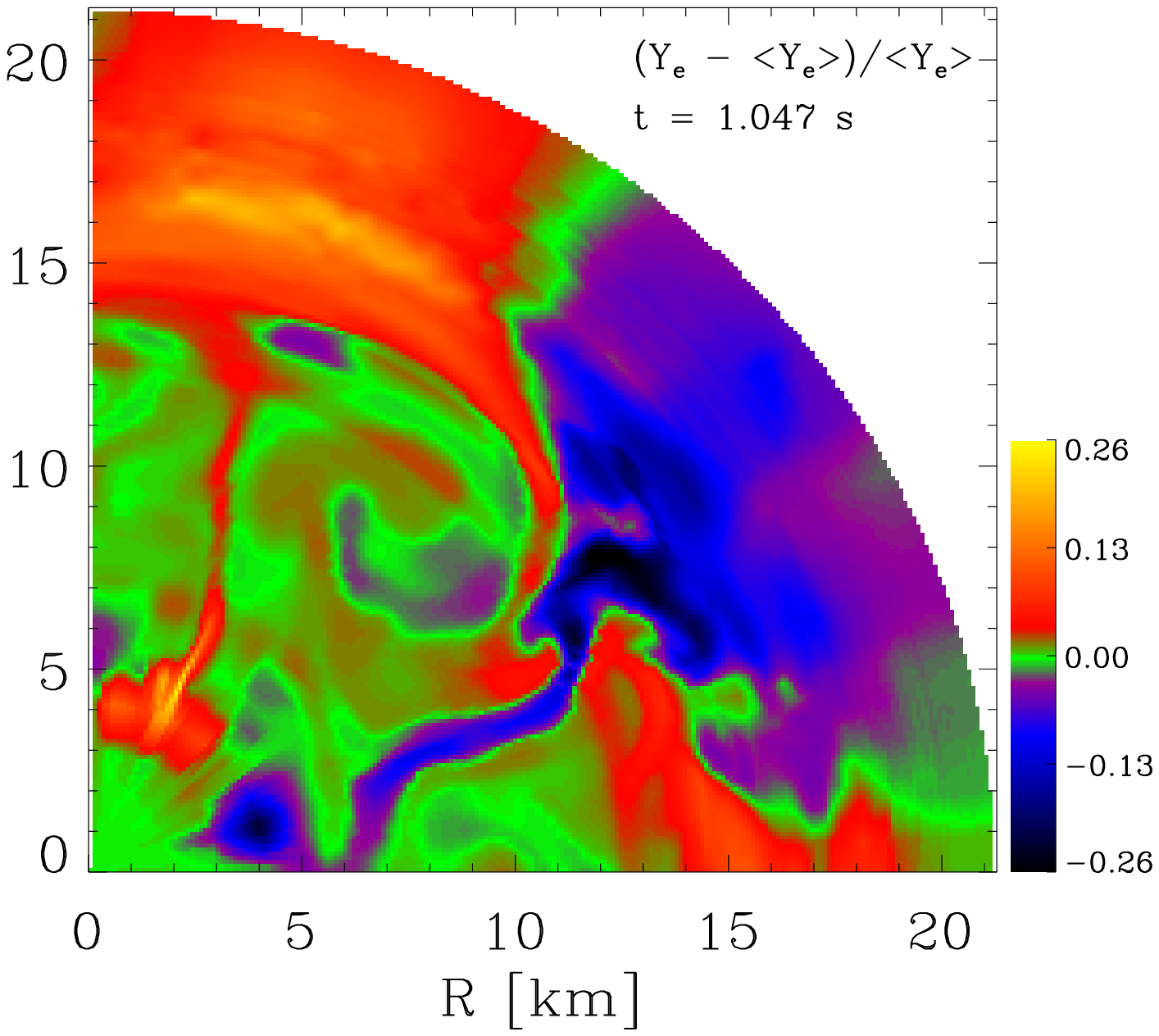}
 \put(-2.0,7.5){{\Large \bf d}}
 \end{tabular}
 \caption[]{
 }
\label{fig:snapshots}
\end{figure*}

The convective pattern is extremely non-stationary and has
most activity on large scales with radial coherence lengths of
several km up to $\sim 10\,{\rm km}$ and convective ``cells''
of 20$^{\circ}$--30$^{\circ}$ angular diameter, at some times
even 45$^{\circ}$ (Fig.~\ref{fig:snapshots}). Significant over-
and undershooting takes place (grey regions
in Fig.~\ref{fig:conv-map}) and the convective mass motions
create pressure waves and perturbations in the
convectively stable NS interior and in the surface layers. The maximum
convective velocities are usually $\sim 4\cdot 10^8\,{\rm cm/s}$, but
peak values of $\sim 10^9\,{\rm cm/s}$ can be reached. These
velocities are typically 5--10\% of the average sound speed
in the star. The kinetic energy of the convection is several
$10^{49}\,{\rm erg}$ at $t\la 1\,{\rm s}$ and climbs to
$\sim 2\cdot 10^{50}\,{\rm erg}$ when the PNS is fully convective.
Relative deviations of $Y_l$ from the angular mean
can be several 10\% (even 100\%) in rising or sinking buoyant
elements, and for $S$ can reach 5\% or more. Rising flows
always have larger $Y_l$ {\it and} $S$ than their surroundings.
Corresponding
temperature and density fluctuations are only $\sim 1$--3\%.
Due to these properties and the problems in applying the Ledoux
criterion with angle-averaged $S$ and $Y_l$
straightforwardly, we suspect that it is hardly possible
to describe the convective activity with a mixing-length treatment
in a 1D simulation.

\begin{figure}
\centering\leavevmode
\epsfxsize=16cm
\epsfbox{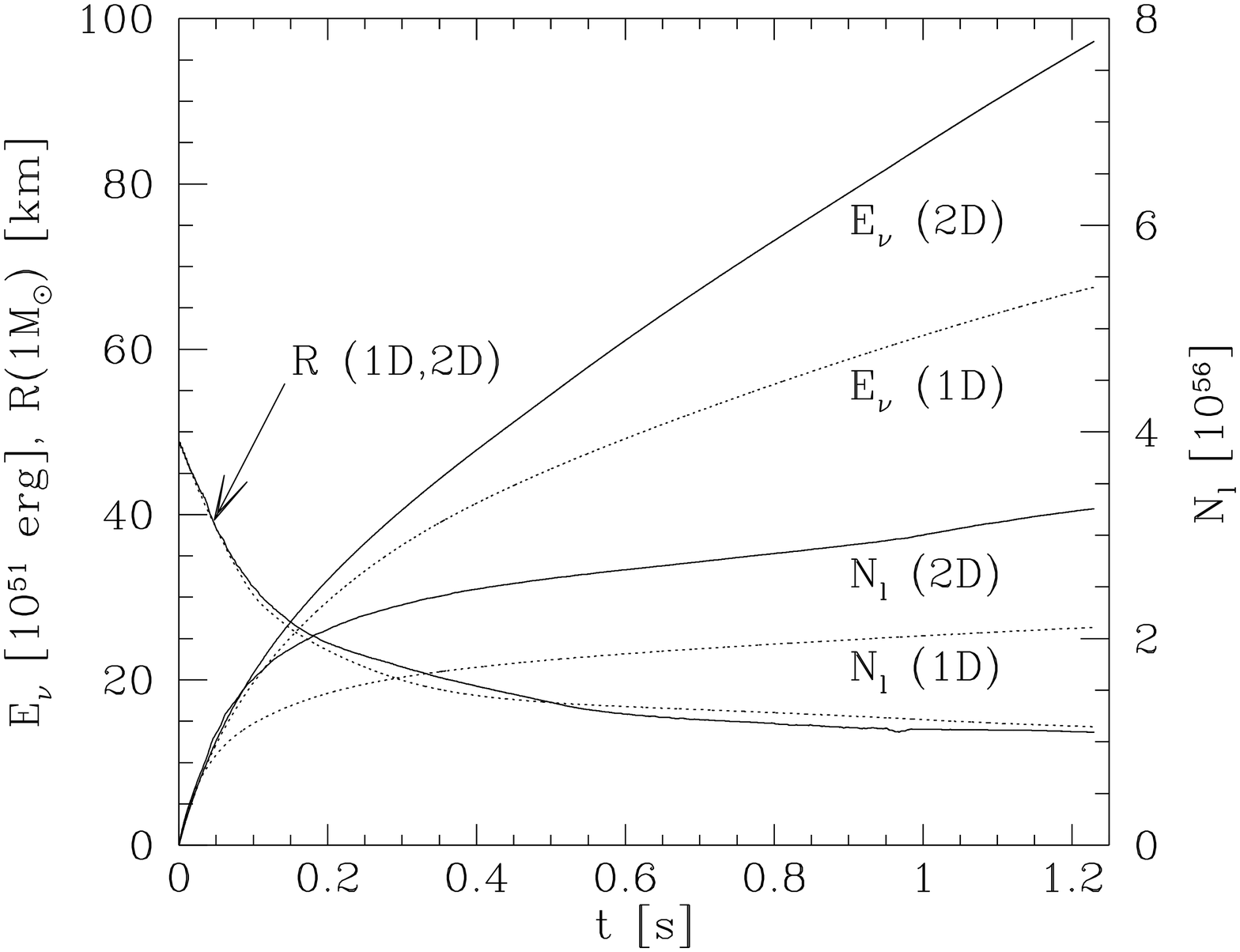}
\caption[]{
}
\label{fig:etot-nlep}
\end{figure}

Our 2D simulation shows that convection in the PNS can encompass
the whole star within $\sim 1\,{\rm s}$ and can continue for at least
as long as the deleptonization takes place, possibly even longer.
A deleptonization ``wave'' associated with the convectively
enhanced transport moves towards the center of the PNS. This
reduces the timescale for the electron fraction $Y_e$ to
approach its minimum central value of about 0.1 from
$\sim 10\,{\rm s}$ in the 1D case, where the lepton loss proceeds
much more gradually and coherently, to only $\sim 1.2\,{\rm s}$
in 2D.
With convection the entropy and temperature near the center rise
correspondingly faster despite a similar contraction of the star
in 1D and 2D
(Fig.~\ref{fig:etot-nlep}). Convection increases the total lepton
number flux and the $\nu$ luminosities by up to a factor of 2
(Fig.~\ref{fig:nu-lum}) and therefore the emitted lepton number
$N_l$ and energy $E_{\nu}$ rise much more rapidly
(Fig.~\ref{fig:etot-nlep}). The convective energy (enthalpy
plus kinetic energy) flux dominates the diffusive $\nu$ energy flux
in the convective mantle after $t \ga 250\,{\rm ms}$ and becomes
more than twice as large later. Since convection takes place somewhat
below the surface, $\nu$'s take over the energy transport
exterior to $\sim 0.9\,M_{\odot}$. Thus the surface $\nu$ flux shows
relative anisotropies of only 3--4\%, in peaks up to $\sim 10\%$,
on angular scales of 10$^{\circ}$--40$^{\circ}$.
Averaged over all directions,
the neutrinospheric temperatures and mean energies
$\ave{\epsilon_{\nu_i}}$ of the emitted $\nu_e$ and $\bar\nu_e$
are higher by 10--20\% (Fig.~\ref{fig:nu-lum}).


\section{Consequences and conclusions\label{sec:conclusions}}

\begin{figure}
\centering\leavevmode
\epsfxsize=16cm
\epsfbox{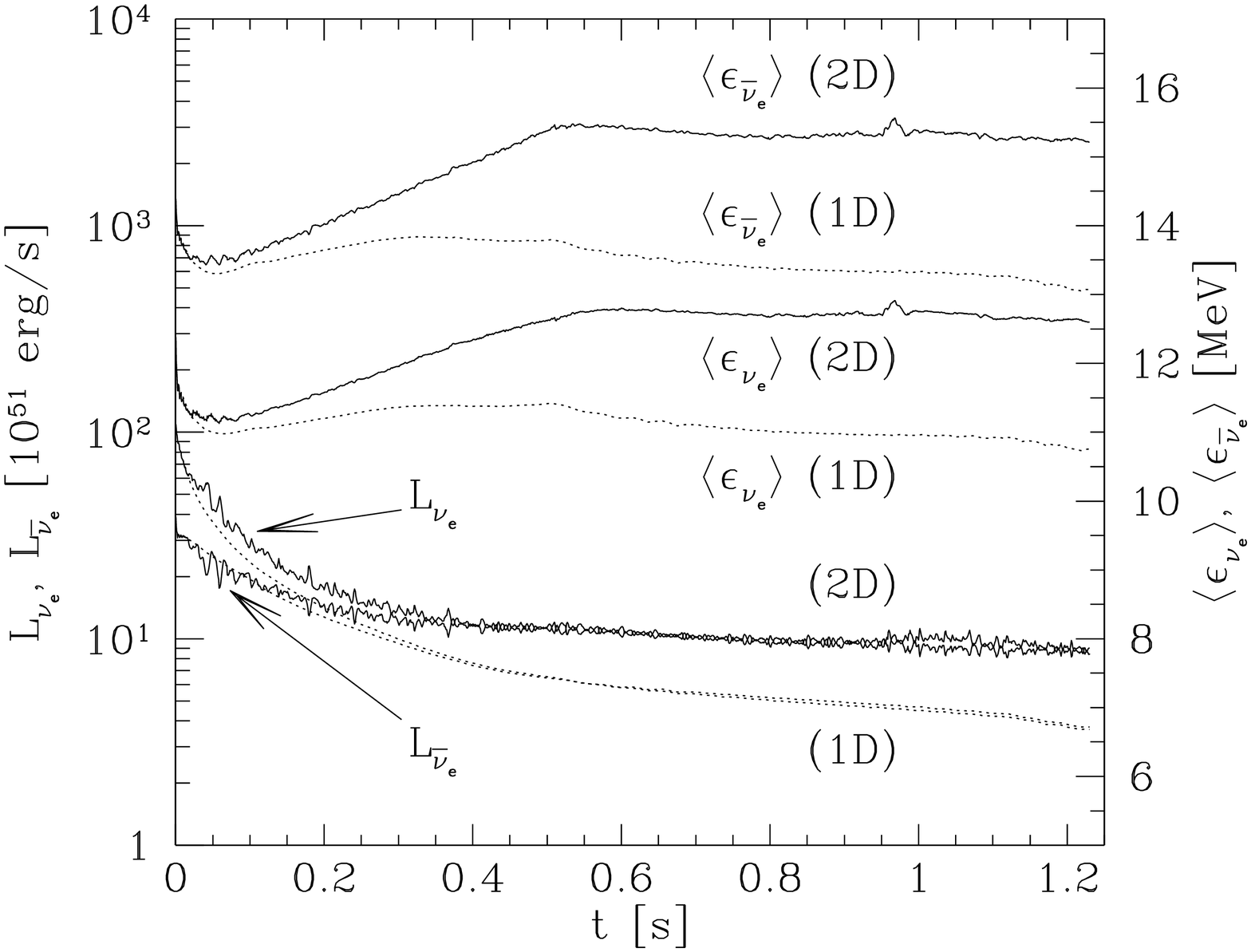}
\caption[]{
}
\label{fig:nu-lum}
\end{figure}

The increase of the $\nu_e$-luminosity
relative to the $\bar\nu_e$-luminosity
during $t \la 0.4\,{\rm s}$ (Fig.~\ref{fig:nu-lum})
will raise $Y_e^{\rm ej}$ in the neutrino-heated SN ejecta.
If weak equilibrium is established,
$\alpha$ particles are absent, and $e^{\pm}$ captures can be ignored,
captures of $\nu_e$ on $n$ and $\bar\nu_e$ on $p$ determine
$Y_e^{\rm ej}\approx
1/\eck{1+(L_{\bar\nu_e}\ave{\epsilon_{\bar\nu_e}})/
(L_{\nu_e}\ave{\epsilon_{\nu_e}})}$ (Qian \& Woosley 1996).
The luminosity ratio enters crucially, since
$\ave{\epsilon_{\bar\nu_e}}/\ave{\epsilon_{\nu_e}} \approx
T_{\bar\nu_e}/T_{\nu_e}$ does not change much due to convection
(compare Fig.~\ref{fig:nu-lum}). The latter fact can be understood
by an analytical neutrino Eddington atmosphere model
(Schinder \& Shapiro 1982) which yields for the temperatures
$T_{\nu_i}$ of the $\nu_i$ energyspheres as functions of the
effective temperature $T_{\rm eff}$:
$T_{\nu_i} = \root 4 \of {\xi_{\nu_i}/2}\,\,T_{\rm eff}\approx
4.6\,\root 4\of {\xi_{\nu_i}{\cal L}_{52}/r_6^2}\,\,{\rm MeV}$,
where
${\cal L}_{52}\equiv (L_{\nu_e}+L_{\bar\nu_e})/
(10^{52}\,{\rm erg/s})$, $r_6\equiv r/(10^6\,{\rm cm})$.
The expression
$\xi_{\nu_i}$ depends on the $p$ and $n$ abundance fractions,
$Y_p$ and $Y_n\approx 1-Y_p$, in the NS atmosphere. For $\nu_e$
one gets $\xi_{\nu_e}\approx 1+1/(Y_n\sqrt{1+0.182Y_p/Y_n})$
and for $\bar\nu_e$ one finds
$\xi_{\bar\nu_e}\approx 1 +1/(Y_p\sqrt{1+0.210Y_n/Y_p})$. The ratio
$T_{\bar\nu_e}/T_{\nu_e} = \root 4\of {\xi_{\bar\nu_e}/\xi_{\nu_e}}$
varies only weakly with the atmospheric composition. Moreover,
one can derive that $Y_e^{\rm ej}$ increases if
$f_{\rm n} > f_{\rm e}^{3/4}$ for
$f_{\rm e}\equiv {\cal L}_{\rm 2D}/{\cal L}_{\rm 1D}$ and
$f_{\rm n}\equiv {\cal N}_{\rm 2D}/{\cal N}_{\rm 1D}$
with ${\cal N}\equiv L_{\nu_e}/\ave{\epsilon_{\nu_e}}
- L_{\bar\nu_e}/\ave{\epsilon_{\bar\nu_e}}$. This is fulfilled
at times $t\la 0.4\,{\rm s}$. The expected increase of
$Y_e^{\rm ej}$ might help to avoid the overproduction
of $N = 50$ nuclei in current SN models (see
Hoffman et al.~1996, McLaughlin et al.~1996).
At times $t\ga 1\,{\rm s}$ the accelerated neutronization
of the PNS will lead to a more rapid increase of 
$\ave{\epsilon_{\bar\nu_e}}$ relative to $\ave{\epsilon_{\nu_e}}$
than in 1D models. This will favor a faster drop of $Y_e^{\rm ej}$
and thus the $n$-rich conditions required for a possible 
r-processing in the neutrino-driven wind.

Future simulations of Ledoux convection in the PNS
that include the progenitor star outside the nascent
NS will have to reveal the effects on the SN explosion mechanism.
PNS models with (baryonic) masses
$M_{\rm ns}\approx 1.5\,...\,1.65\,M_{\odot}$ must be considered
to investigate implications for the $\nu$ signal detected from
SN~1987A. A more elaborate description of the $\nu$ transport
and the use of different EOSs are also required. Convection in
the PNS influences the structure of NS magnetic fields
(Thompson \& Duncan 1993), produces gravitational wave emission,
and can cause NS recoils by anisotropic $\nu$ emission
(Thompson \& Duncan 1993). For non-stationary convection
with typical coherence lengths $l_{\rm c}\sim 10^5\,{\rm cm}$ and
overturn timescales $t_{\rm o}\la R_{\rm ns}/v_{\rm c}
\la 10\,{\rm ms}$, one estimates a stochastic anisotropy of
$\alpha\sim (l_{\rm c}/R_{\rm ns})\sqrt{t_{\rm o}/t_{\rm ns}}
\la 10^{-2}$ ($t_{\rm ns}$: NS cooling timescale) which leads
to kick velocities $v_{\rm ns}\approx \alpha E_{\nu}/(M_{\rm ns}c)$
of a few $100\,{\rm km/s}$, dependent on the energy
$E_{\nu}\la {3\over 5}GM_{\rm ns}^2/R_{\rm ns}$
emitted anisotropically in neutrinos.

\bigskip

We would like to thank S.W.~Bruenn for kindly providing us with
the post-collapse model to be used as initial model in our
simulations. 
This work was supported
by the Sonderforschungsbereich 375-95 for
Astro-Particle Physics of the Deutsche Forschungsgemeinschaft.
The computations were
performed on the Cray-YMP 4/64 and the Cray-EL98~4/256
of the Rechenzentrum Garching.



\refs

Bethe H.A., Brown G.E., Cooperstein J., 1987, ApJ~332, 201

Bruenn S.W., 1993, in {\it Nuclear Physics in the Universe},
   eds. M.W. Guidry and M.R. Strayer, IOP, Bristol, p.~31

Bruenn S.W., Mezzacappa A., 1994, ApJ~433, L45

Bruenn S.W., Mezzacappa A., Dineva T., 1995, Phys. Rep. 256, 69

Bruenn S.W., Dineva T., 1996, ApJ~458, L71

Burrows A., 1987, ApJ~318, L57

Burrows A., Hayes J., Fryxell B.A., 1995, ApJ~450, 830

Burrows A., Lattimer J.M., 1986, ApJ~307, 178

Burrows A., Lattimer J.M., 1988, Phys.~Rep.~163, 51

Colella P., Woodward P.R., 1984, J.~Comp.~Phys.~54, 174

Epstein R.I., 1979, MNRAS~188, 305

Fryxell B.A., M\"uller E., Arnett W.D., 1989,
   MPA-Preprint 449, Garching

Hillebrandt W., 1987, in {\it High Energy Phenomena
   around Collapsed Stars},
   ed.~F.~Pacini, Reidel, Dordrecht, p.~73

Herant M., Benz W., Hix W.R., Fryer C.L., Colgate S.A.,
   1994, ApJ~435, 339

Hoffman R.D., Woosley S.E., Fuller G.M., Meyer B.S.,
   1996, ApJ~460, 478

van den Horn L.J., van Weert C.G., 1981, ApJ~251, L97

Janka H.-Th., M\"uller E., 1996, A\&A~306, 167

Keil W., 1996, PhD Thesis, TU M\"unchen

Keil W., Janka H.-Th., 1995, A\&A~296, 145

Lattimer J.M., Swesty F.D., 1991, Nucl.~Phys.~A535, 331

Mayle R.W., Wilson J.R., 1988, ApJ~334, 909

Mezzacappa A., Calder A.C., Bruenn S.W., Blondin J.M.,
   Guidry M.W., Strayer M.R., Umar A.S., 1996, preprint,
   submitted to ApJ Letters

McLaughlin G.C., Fuller G.M., Wilson J.R., 1996, ApJ, in press

Porter D.H., Woodward P.R., 1994, ApJS~93, 309

Qian Y.-Z., Woosley S.E., 1996, ApJ, in press

Schinder P.J., Shapiro S.L., 1982, ApJ~259, 311

Sumiyoshi K., Suzuki H., Toki H., 1995, A\&A~303, 475

Thompson C., Duncan R.C., 1993, ApJ~408, 194

Wilson J.R., Mayle R.W., 1988, Phys.~Rep.~163, 63

Wilson J.R., Mayle R.W., 1989, in {\it The Nuclear
   Equation of State, Part~A}, eds.~W.~Greiner and
   H.~St\"ocker, Plenum Press, New York, p.~731

Wilson J.R., Mayle R.W., 1993, Phys.~Rep.~227, 97

Woosley S.E., Pinto P.A., Ensman L., 1988, ApJ~324, 466
 
\endrefs
                                                                          
\vfill\eject
%
%
%
%
\centerline{{\top Figure captions:}}
 
\bigskip\medskip
\noindent
FIG.~1.
Convective (baryon) mass region inside the PNS vs.~time
for the 2D simulation. Black indicates regions which are Ledoux
unstable or only marginally stable, grey denotes over- and
undershooting regions where $|v_{\theta}| > 10^7\,{\rm cm/s}$.
\hfill\break
 
\bigskip\noindent
FIG.~2.
Angle-averaged $S$ and $Y_l$ profiles in the PNS.
Thick solid lines indicate regions
that are unstable or only marginally stable against Ledoux convection,
crosses mark boundaries of over- and
undershooting regions where $|v_{\theta}| > 10^7\,{\rm cm/s}$.
\hfill\break
 
\bigskip\noindent
FIG.~3.
Panels {\bf a} and {\bf b} show the
absolute values of the velocity for the 2D simulation at times
$t = 0.525~{\rm s}$ and $t = 1.047~{\rm s}$, respectively,
color-coded in units of $10^8\,{\rm cm/s}$.
The computation was performed in an angular wedge of $90^{\circ}$
between $+45^{\circ}$ and $-45^{\circ}$ around the equatorial
plane. The PNS has contracted to a radius of about
$21\,{\rm km}$ at the given times.
Panels {\bf c} and {\bf d} display the relative deviations of
the electron fraction $Y_e$ from the angular means $\ave{Y_e}$
at each radius for the same two instants. The maximum deviations are
of the order of 30\%. Lepton-rich matter rises while deleptonized
material sinks in. Comparison of both times shows that the
inner edge of the convective layer moves inward from about
$8.5\,{\rm km}$ at $t = 0.525~{\rm s}$ to less than
$2\,{\rm km}$ at $t = 1.047~{\rm s}$.
\hfill\break
 
\bigskip\noindent
FIG.~4.
Radius of the $M = 1\,M_{\odot}$ mass shell and
total lepton number $N_{\rm l}$ and energy
$E_{\nu}$ radiated away by $\nu$'s vs.~time for
the 2D (solid) and 1D (dotted) simulations.
\hfill\break
 
\bigskip\noindent
FIG.~5.
$\nu_e$ and $\bar\nu_e$ luminosities and
mean energies vs.~time for the 2D
simulation (solid) compared with the 1D
run (dotted).
%
%
%
%
%
%
%
%
%
%
%
%
%
%
%
%
%
%
%
%
%
%
%
%
%
%
%

\end{document}